\documentclass[11pt]{article}

\usepackage[preprint]{acl}

\usepackage{times}
\usepackage{latexsym}
\usepackage{amsmath}
\usepackage{amssymb}  
\usepackage{xspace}
\usepackage{multirow}
\usepackage{enumitem}
\usepackage{algorithm2e}
\usepackage{algorithmicx}
\usepackage{algpseudocode}
\usepackage{threeparttable}
\usepackage{booktabs}
\usepackage{svg}
\usepackage[most,skins]{tcolorbox}
\usepackage{xcolor}      
\usepackage{makecell} 
\usepackage{xcolor}
\usepackage{amsmath}     
\usepackage{subcaption} 
\usepackage{graphicx}
\usepackage{url}

\definecolor{DarkGreen}{RGB}{0,100,0} 

\newcommand{\sys}{SeCuRepair\xspace}

\usepackage[T1]{fontenc}
\usepackage{enumitem}

\usepackage[utf8]{inputenc}

\usepackage{microtype}

\usepackage{inconsolata}

\usepackage{graphicx}

\title{SeCuRepair: Semantics-Aligned, Curriculum-Driven, and Reasoning-Enhanced Vulnerability Repair Framework}

\author{
  \textbf{Chengran Yang}$^{1}$ \quad
  \textbf{Ting Zhang}$^{2}$ \quad
  \textbf{Jinfeng Jiang}$^{1}$ \quad
  \textbf{Xin Zhou}$^{1}$ \quad
  \textbf{Haoye Tian}$^{3}$ \quad
  \textbf{Mingzhe Du}$^{4}$ \\  
  \textbf{Jieke Shi}$^{1}$ \quad
  \textbf{Junkai Chen}$^{1}$ \quad
  \textbf{Yikun Li}$^{1}$ \quad
  \textbf{Eng Lieh Ouh}$^{1}$ \quad
  \textbf{Lwin Khin Shar}$^{1}$ \quad
  \textbf{David Lo}$^{1}$ \\
  \\
  $^{1}$Singapore Management University, Singapore \quad
  $^{2}$Monash University, Australia \\
  $^{3}$Aalto University, Finland \quad
  $^{4}$National University of Singapore, Singapore\\
}

\begin{document}
\maketitle
\begin{abstract}
The rapid accumulation of software vulnerabilities has outpaced manual remediation, creating an urgent need for Automated Vulnerability Repair (AVR). However, existing methods suffer from syntactic overfitting, mimicking surface forms without understanding the underlying repair logic, and fail to generalize to complex fixes. 
To transcend these limitations, we propose SeCuRepair, a reliable, scalable, and efficient RL-based AVR framework. 
By introducing a semantic-aware reward, SeCuRepair optimizes for code semantic equivalence rather than lexical mimicry. Furthermore, SeCuRepair incorporates an expert-aligned reasoning mechanism that explicitly grounds patch generation in a structured diagnosis.
Finally, SeCuRepair introduces a difficulty-based curriculum that progressively disentangles the optimization barriers of entangled multi-hunk repairs.
Extensive evaluations on a rigorous repository-level split demonstrate that SeCuRepair significantly outperforms state-of-the-art baselines via automatic evaluation and human study. 

\end{abstract}

\section{Introduction}

The rapid accumulation of software vulnerabilities has overwhelmed manual remediation capacities. The National Vulnerability Database~(NVD) published 49,230 Common Vulnerabilities and Exposures~(CVE) records in 2025 alone, yet due to slow human remediation, an overall total of 27,900 CVEs remain unanalyzed~\cite{nvd_dashboard}. With manual fixes often requiring weeks ~\cite{cve_fix_time2}, there is a critical need for effective Automatic Vulnerability Repair~(AVR).

To address this backlog, existing approaches rely on Supervised Fine-Tuning (SFT)~\cite{vrepair, vulmaster, favor}, training models to map vulnerable code to a specific human-written patch. However, this formulation is fundamentally misaligned with the intrinsic flexibility of AVR: real-world vulnerability admits diverse semantically equivalent fixes, while SFT optimizes for the exact reconstruction of a unique syntactic solution. This rigid objective penalizes generating semantically equivalent but syntactically different patches, forcing models to overfit to superficial coding patterns.

Reinforcement learning~(RL) transcends SFT's rigidity by shifting the optimization objective from token-level imitation to outcome-driven exploration~\cite{shao2024deepseekmath}. However, applying RL to AVR is impeded by two fundamental challenges: \emph{the lack of a reliable reward signal and the complexity of the exploration landscape.} 

\textbf{First, constructing a reliable reward signal for AVR is difficult.}
While execution-based rewards are standard in Code LLMs~\cite{guo2025deepseek, hui2024qwen2}, real-world AVR datasets lack the large-scale build environments and test cases required for dynamic verification.
Consequently, execution-free alternatives prove inadequate in AVR: 
1) Lexical Rewards~(e.g., SWE-RL~\cite{wei2025swe}) rely on string similarity to the reference, which effectively reintroduces SFT’s rigidity by penalizing valid but syntactically different patches; 
2) LLM-as-Judge is unreliable in AVR, often failing to distinguish between vulnerable and patched functions~\cite{primevul}, thereby providing noisy and indiscriminative signals. This leaves a critical gap for a reward design that is execution-free yet rewards semantic equivalence.

\textbf{Second, efficient exploration is hindered by the inherent complexity of vulnerability repair.} 
Direct RL on AVR often yields sub-optimal policies due to two critical characteristics: 
1) \emph{Reasoning-dependent Nature}: effective repair is an inherent chain-of-thought process, requiring the inference of underlying vulnerability logic and analyzing of the root cause before generating a patch. Bypassing this reasoning process leads to a reasoning gap, rendering the exploration phase unstable.
2) \emph{Entangled Learning Objectives}: complex repairs spanning multiple code hunks (i.e., non-contiguous edited regions in a function) require the model to simultaneously master precise local edits (for security) and global structural synchronization (for syntactic validity). Attempting to optimize these entangled objectives concurrently overwhelms the exploration process, often trapping the LLM exploration in local optima.

To bridge these gaps, we propose SeCuRepair, which enables reliable, scalable, and efficient RL-based vulnerability repair. 
To bridge the gap of reliable reward, we propose a \textbf{multi-grained reward}. We proxy patch quality through a composite metric of lexical~(BLEU), syntactic~(AST), and semantic~(Data and Control Flow Graph) similarity to human reference. 
This formulation anchors the reward in semantic invariants, rewarding semantically equivalent but syntactically diverse patches.
Meanwhile, it enables scalable training across large-scale, real-world AVR datasets where build environments are missing.

For efficient exploration, we propose a \textbf{reasoning-enhanced initialization} followed by a \textbf{curriculum training strategy}.
Specifically, we employ an expert-aligned reasoning protocol to initialize the policy, ensuring LLM executes a diagnostic process before attempt to fix.
Meanwhile, we organize training samples by complexity, utilizing the number of code hunks as a proxy. This curriculum guides the LLM to first master isolated local fixes by starting with single-hunk samples, before progressively advancing to tackle global structural coordination in complex scenarios.

We evaluate the efficacy of SeCuRepair on established benchmark BigVul~\cite{fan2020ac} and our newly proposed PrimeVul\textsubscript{AVR}, which employing a strict repository-level splitting strategy to avoid data leakage.
SeCuRepair demonstrates significant superiority over state-of-the-art baselines, achieving a 37.21\% improvement in CodeBLEU on BigVul and 33.58\% on PrimeVul\textsubscript{AVR}. Human evaluation further confirms \sys performs on par with GPT-4o.

In summary, our main contributions are as follows: 1) We propose SeCuRepair, an RL-based framework that transcends SFT rigidity via a semantic-aware reward. 2) We devise a curriculum training strategy combined with reasoning-enhanced distillation for efficient exploration.
3) We establish a rigorous cross-repository evaluation protocol to eliminate the data leakage inherent in random splits, and introduce a new AVR benchmark PrimeVul\textsubscript{AVR}.

\section{Related Work}
\label{sec:related_work}

\noindent{\bf AVR approaches.} 
Rule-based methods, ranging from template-guided~\cite{lin2007autopag} to constraint-based solvers~\cite{zhang2022program,gao2021beyond}. They rely on either rigid, manually defined patterns or expensive runtime execution, making them ineffective for the majority of vulnerabilities where build environments are unavailable.
Leading learning-based approaches such as VulMaster~\cite{vulmaster} and FAVOR~\cite{favor} improve repair performance by incorporating external knowledge and structural context based on the SFT objective. Yet, their reliance on SFT inherently encourages lexical mimicry over semantic understanding. 

\noindent{\bf RL in software engineering.}
Reinforcement Learning has shown promise in various software engineering tasks~\cite{guo2025deepseek, hui2024qwen2, wei2025swe, yang2024acecode}, typically relying on execution feedback as reward. However, this is impractical for AVR due to the scarcity of large-scale build environments.
Existing execution-free alternatives also fall short.
Lexical rewards (e.g., SWE-RL~\cite{wei2025swe}) measure string similarity to a reference, but this rigid metric penalizes syntactically diverse yet semantically equivalent solutions, providing a biased signal.
Model-based rewards~\cite{dutta2024applying, sghaier2025leveraging} employ LLMs as judges, but these prove unreliable for security tasks, as even SOTA models struggle to distinguish between vulnerable code and its patched version~\cite{primevul}, which is the core of AVR rewarding.
In contrast, our approach bridges this gap by introducing a multi-grained, semantic-aware reward, proxying patch quality through static analysis of syntactic and semantic equivalence.

\section{Problem Statement and Motivation}
\label{sec:motivation}
\begin{figure*}[t]
\centering
\vspace{-3mm}

\begin{minipage}[b]{0.49\textwidth}
  \centering
  \includegraphics[width=\linewidth]{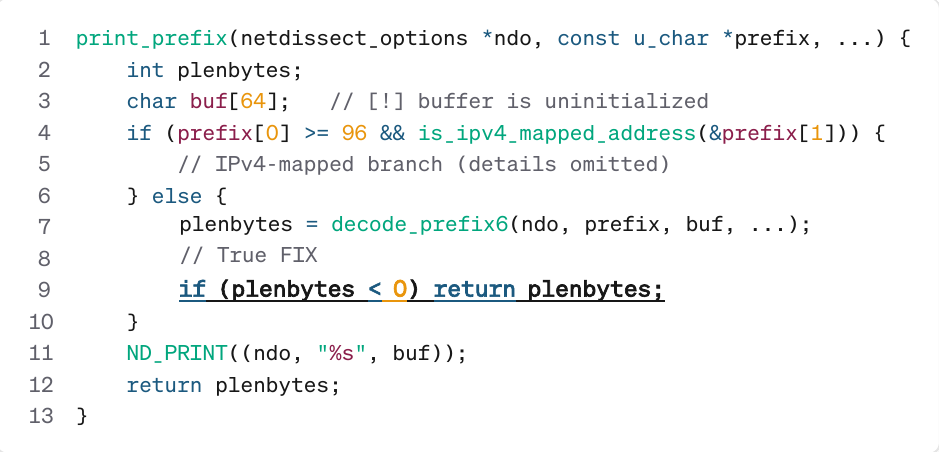}
\end{minipage}
\hfill
\begin{minipage}[b]{0.49\textwidth}
  \centering
  \includegraphics[width=\linewidth]{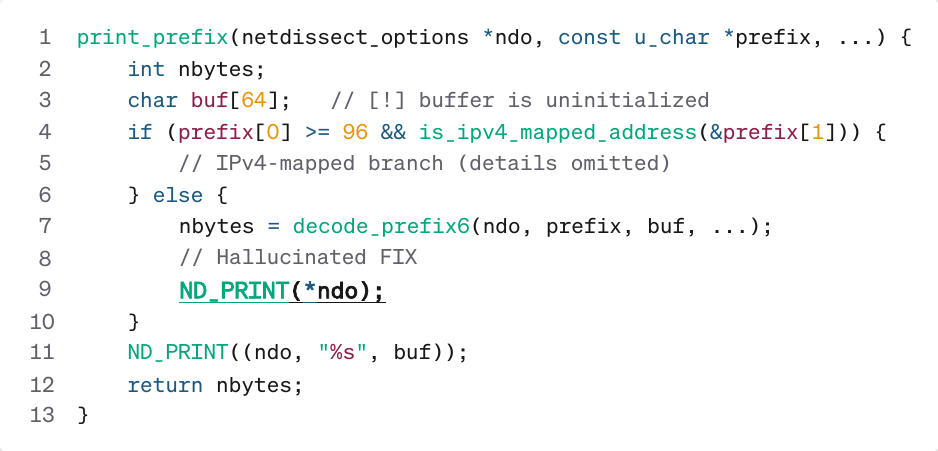}
\end{minipage}

\vspace{-0.2cm}
\caption{
Syntactic overfitting of SFT: The SFT model generates a correct fix for the buffer over-read guard with the original input, but fails to generalize when a local variable is simply renamed from \textsc{plenbytes} to \textsc{nbytes}.
}
\label{fig:repo_gap_qualitative}
\vspace{-4mm}
\end{figure*}

In this section, we formulate the task of AVR and analyze the limitations of existing AVR approaches: 1) Syntactic Overfitting; 2) Suboptimization on Complex Repair; and 3) Biased Evaluation Setting.

\subsection{Problem Definition}
Following prior works~\cite{vrepair,vulmaster,favor}, we formulate AVR as a function-level sequence-to-sequence generation task.
Given a vulnerable function $X$ and auxiliary information $L$ (e.g., information of vulnerability localization and Common Weakness Enumeration), the goal is to generate a patched function $Y$ that fixes the vulnerability in $X$. 

\subsection{Syntactic Overfitting of SFT in AVR}
\label{sec:Syntactic_Overfitting}
To probe whether existing SFT approaches~\cite{wang2021codet5,vulmaster, favor} learn semantic repair patterns or overfit to syntactic cues, we follow~\cite{rahman2024causal} and test their performance under a simple refactoring: renaming local variables of input while preserving code semantics (see Appendix~\ref{app: pertubation} for details).
We observe that all SFT models are sensitive to these minor changes, failing to generate valid patches in 40–70\% of cases.
Figure~\ref{fig:repo_gap_qualitative} provides an example of this failure. 
A simple variable renaming (\texttt{plenbytes} $\to$ \texttt{nbytes}) causes VulMaster to regress from a correct fix to an irrelevant hallucination.
This provides empirical evidence that SFT models heavily rely on superficial variable correlations.
In this paper, we propose an RL training framework that rewards semantically equivalent patches to mitigate syntactic overfitting.

\subsection{Suboptimization on Multi-hunk Repair}
\label{sec:repo_gap}
A critical oversight in current AVR research is the lack of optimization for complex, multi-hunk repair. 
Real-world fixes often require changes across multiple discontinuous code regions, yet existing methods treat all samples uniformly.
To quantify this impact, we stratified the performance of representative baselines, VulMaster~\cite{vulmaster} and FAVOR~\cite{favor}, by the number of repair hunks in human reference.
Performance drops with complexity: even moving from 1 to 2 hunks causes 11\% to 20\% CodeBLEU drops across all approaches (see Figure~\ref{fig:multi_hunk_repair} for full results). 

Our manual inspection reveals that 55\% of multi-hunk failures stem from global inconsistencies. We attribute this failure to entangled learning objectives: the model faces the dual burden of executing precise local security edits while simultaneously maintaining global syntactic synchronization. 
We propose a curriculum-training strategy to mitigate this issue, which guides the LLM to first master local fixes by starting with single-hunk samples.
This setting smooths the optimization landscape, allowing the model to learn patch patterns without global interference~\cite{curriculumlearning}. 
We then introduce multi-hunk samples, allowing LLM to focus more on aligning global dependencies.

\subsection{Data Leakage in Conventional Evaluation} 
Existing AVR approaches typically rely on function-level random splitting~\cite{vulmaster, favor}. However, this setting poses a significant data leakage risk~\cite{zhang2025benchmarking}, as it allows models to exploit project-specific patterns or "peek" at future fixes within the same repository. We quantify this threat by comparing model performance under the random-split versus a strict cross-repository protocol, where all functions from a specific project are confined to a single split (train/val/test). The results reveal a severe generalization gap. Under the strict repository-level split, performance degrades sharply across all baselines: CodeBLEU scores drop by a relative 21.49\%--29.70\%, while exact match scores plummet by 87.88\% to 89.86\% (see Appendix~\ref{app:repo_gap} for full details). These findings highlight the leakage issue in a random-split setting, inspiring us to use a cross-repository setting in the following sections.

\section{Approach}

Driven by the above limitations, we propose \sys, an RL-based AVR framework to enforce LLM to learn the underlying logic of vulnerability repair.
The design of \sys is guided by three core principles:

    \noindent \textbf{1) Reasoning-enhanced Initialization:} 
    Vulnerability repair inherently demands complex logic: root cause analysis, vulnerability type classification, and devising a fix strategy. To mitigate the optimization instability caused by this reasoning gap, we warm-start the model via knowledge distillation, leveraging a structured protocol that simulates a security-expert cognitive workflow.
    
    \noindent \textbf{2) Semantics-Aware Optimization:}
    To overcome the syntactic overfitting of SFT, we propose a multi-grained static reward signal for patch quality, which enables the measurement of semantic equivalence.
    
    \noindent \textbf{3) Difficulty-based Curriculum Learning :}
    To dismantle the optimization barrier in complex multi-hunk repairs, we employ a difficulty-based curriculum.
    This approach decouples the learning process by progressively increasing task complexity, allowing the model to master local security edits before tackling global synchronization.

\begin{figure*}[t]
  \centering
  \includegraphics[width=0.9\textwidth]{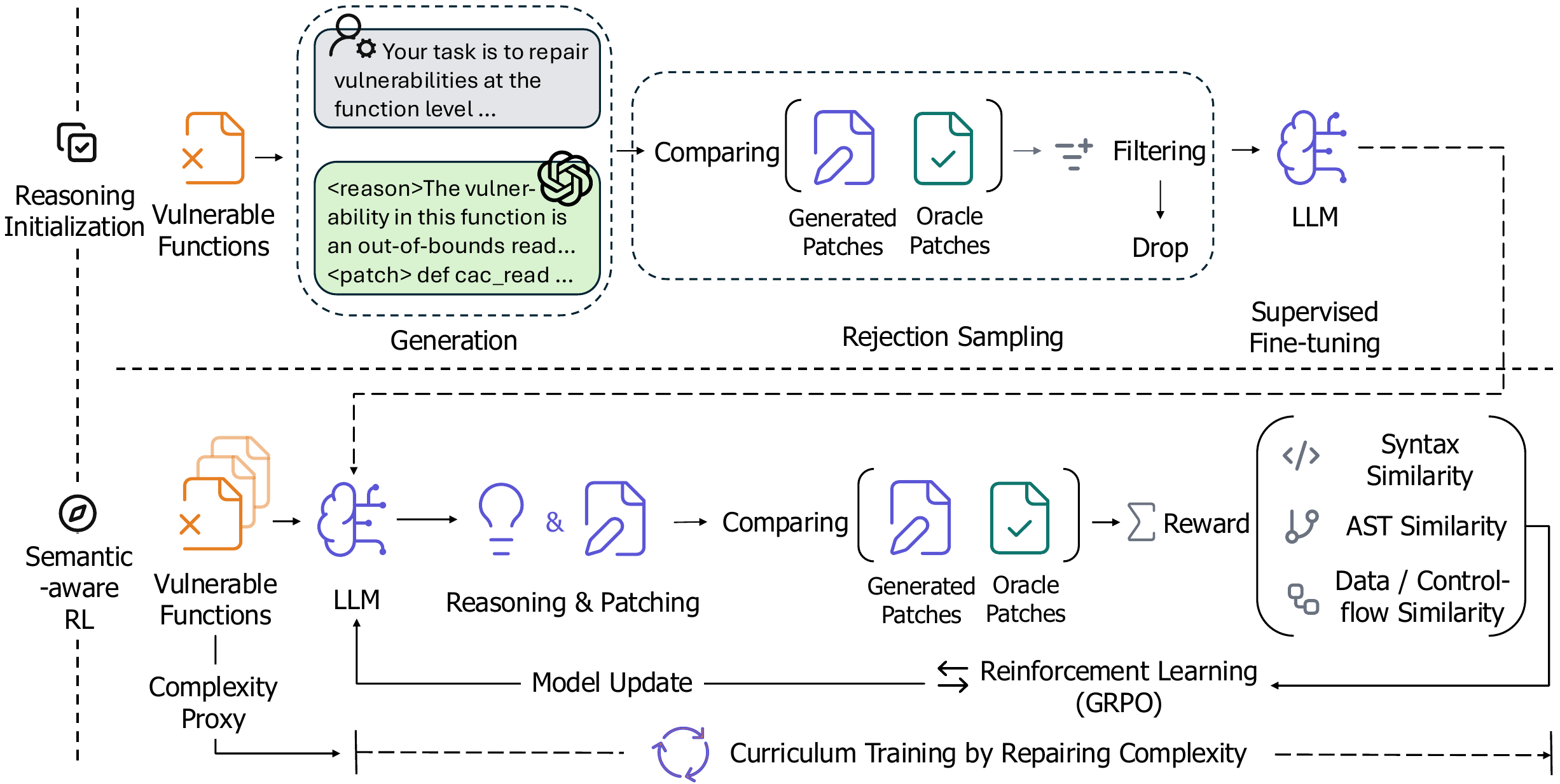}
  \vspace{-0.2cm}
  \caption{Overview of the \sys training pipeline. (1) Stage 1: Reasoning-enhanced Initialization. A teacher LLM generates (reasoning, patch) candidates, which are filtered via rejection sampling to create a high-quality dataset. The student model is then fine-tuned to learn the reason-then-edit format. (2) Stage 2: Semantics-Guided RL. The SFT model is refined using RL. The policy is rewarded based on a composite score of lexical (BLEU), syntactic (AST), and sematic (CFG \& DFG) similarity to the oracle. Stage 2 is organized with a curriculum, progressively training the model on repairs with an increasing number of hunks to master multi-hunk fixes.} 
  \label{fig:framework}
  \vspace{-5mm}
\end{figure*}

These principles are realized through a two-stage training pipeline, as shown in Figure~\ref{fig:framework}.
In the first stage, we bootstrap the model's reasoning ability. 
We employ a commercial teacher LLM to generate high-quality (reasoning, patch) examples with heuristic-based rejection sampling, and then perform SFT on the student model.
In the second stage, we refine the model's patching ability using GRPO algorithm~\cite{shao2024deepseekmath}. 
The model is optimized through a semantic-aware RL optimization, which uses rewards measuring lexical, syntactic, and semantic correctness, and is guided by difficulty-based curriculum training, which gradually increases task complexity to multi-hunk fixes.

\subsection{Reasoning-enhanced Initialization}

The first stage of our pipeline aims to initialize the model to follow a reason-then-edit paradigm through knowledge distillation. 

\noindent \textbf{Knowledge Distillation.}
To achieve this, the first step is to construct a dataset of (reasoning, patch) pairs by distilling reasoning knowledge from a teacher LLM (i.e., GPT-5 mini).
To fully exploit the reasoning capabilities of teacher LLMs, we define a structured reasoning protocol that mirrors the cognitive workflow of human security experts: (1) Vulnerability Classification: characterize the CWE type and security impact; (2) Root Cause Analysis: trace the logical flaw triggering the vulnerability; and (3) Repair Planning: formulate a high-level repairing strategy before patching. Extracted reasoning chains that adhere to this protocol effectively transfer the teacher's security diagnostic capabilities to the student model.


\noindent \textbf{Rejection Sampling.}
While the expert workflow guides the reasoning process, it does not guarantee correctness. LLMs often produce plausible-sounding but functionally incorrect patches~\cite{favor}.
Therefore, we adopt a two-step rejection-sampling strategy to denoise reasoning data: 1) we apply syntactic filtering to discard responses that violate the output schema; 2) we apply semantic filtering to retain data only if its patch is similar to the human-reference patch using CodeBLEU as a heuristic proxy.

\noindent \textbf{SFT.}
After rejection sampling, we fine-tune the student model with a standard SFT objective.
The resulting model is capable of generating reasoning before patching and serves as the initial policy for the subsequent reinforcement learning stage.




\subsection{Semantics-Aware RL}\label{sec:RL_Stage}

Following SFT stage, \sys employs RL with semantics-aware rewards, enabling the model to explore the solution space and reward semantic equivalent solutions over rigid textual identity.

We employ an on-policy RL algorithm GRPO~\cite{shao2024deepseekmath} for this stage. 
During training, the model generates a (reasoning, patch) pair for a given vulnerability. We then compute a scalar reward $Re$ based on the generated patch against its human reference.
This semantics-aware reward moves beyond token overlap to measure the lexical (BLEU), syntactic (Abstract Syntax Tree, AST), and semantic ( DFG and CFG) similarities. 
It is then used to update the model's policy, encouraging it to generate patches that achieve higher semantic agreement with human reference.


\subsubsection{Reward Design}
\label{sec:rewarding}

Our reward assesses the agreement between generated patch and the reference at three granularities: 

\vspace{0.1cm}
\noindent \textbf{Lexical Agreement.}
We calculate BLEU score between the generated and reference patches, which measures the proportion of overlapping $n$-grams.
It introduces a token-level reward signal that incentivizes the use of correct keywords and identifiers, thereby suppressing token-level hallucinations.
BLEU score for a generated patch $\hat{y}$ against an oracle patch $y$ is defined as:
$\text{BLEU} = \text{BP} \cdot \exp \left( \sum_{n=1}^{N} \tfrac{\log p_n}{N}  \right),$
where $p_n$ is the $n$-gram precision, 
and $\text{BP}$ is the brevity penalty to discourage excessively short generations. 

\vspace{0.1cm}
\noindent\textbf{Syntactic Agreement.}
AST similarity evaluates the structural alignment between the generated patch and the oracle.
Each code snippet is decomposed into a set of subtrees, and similarity is computed as the proportion of matched subtrees. 
Formally, given $S(\hat{y})$ and $S(y)$ as the sets of subtrees from a generated patch $\hat{y}$ and an
oracle patch $y$ respectively, we define AST similarity as: $\text{Sim}_{\text{AST}}(\hat{y}, y) = \frac{|\,S(\hat{y}) \cap S(y)\,|}{|\,S(y)\,|}$.

\vspace{0.1cm}
\noindent \textbf{Semantic Agreement.}
We measure semantic consistency from complementary perspectives: CFG (control-flow graph) and DFG (data-flow graph) similarity.
The DFG represents how values propagate through a program. 
Each code snippet is decomposed into a set of data-dependency tuples $(v_i, v_j)$. 
We measure similarity as the fraction of matched edges: $\text{Sim}_{\text{DFG}}(\hat{y}, y) = \frac{|\,E(\hat{y}) \cap E(y)\,|}{|\,E(y)\,|}$,
where $E(\hat{y})$ and $E(y)$ denote the sets of data-flow edges extracted from $\hat{y}$ and $y$, respectively. 

Complementarily, the CFG captures the algorithmic structure and execution logic. 
We extract CFGs using the Joern parser, where nodes represent statements and edges denote control transfers.
However, computing precise structural similarity metrics, such as the Graph Edit Distance (GED) with structural matching (i.e., ignoring variable naming differences), is an NP-hard.  
This computational complexity makes GED inappropriate for online RL, which requires instant reward feedback. 
To address this efficiency bottleneck, we employ the Weisfeiler-Lehman (WL) Graph Kernel~\cite{shervashidze2011weisfeiler} as a fast, polynomial-time approximation.
The WL kernel maps each graph to a high-dimensional feature vector $\phi(G)$ by iteratively aggregating neighborhood labels to capture topological sub-structures (e.g., chains, cycles). 
We define the CFG similarity as the normalized kernel value:
$\text{Sim}_{\text{CFG}}(\hat{y}, y) = \frac{k_{\text{WL}}(\hat{y}, y)}{\sqrt{k_{\text{WL}}(\hat{y}, \hat{y}) \cdot k_{\text{WL}}(y, y)}}$.


\vspace{0.2cm}
\noindent \textbf{Final Reward.}
We first define a syntactic check to ensure patches are structurally valid and can be parsed into AST via Tree-sitter; failed patches receive a zero reward.
We then define a reward vector $\mathbf{r} = \langle \text{BLEU}, \text{Sim}_{\text{AST}},$ $\text{Sim}_{\text{DFG}}, \text{Sim}_{\text{CFG}} \rangle$ for valid fix, where the components represent these agreement scores.
We normalize these scores into a single reward ranging from [0, 1] by taking their mean: $Re(\hat{y}, y) = \frac{\|\mathbf{r}\|_1}{n}$,
where $\|\mathbf{r}\|_1$ is the L1-norm of the vector, representing the sum of the absolute values of its components. 
This balanced reward function encourages the model to generate patches that are lexically faithful, syntactically consistent, and semantically coherent with the oracle.

\subsection{GRPO Optimization}
In the RL stage, we optimize the policy model using Group Relative Policy Optimization (GRPO)~\cite{shao2024deepseekmath} with the reward signal $Re(\hat{y}, y)$. 
Let $r_{ij}(\theta) = \frac{\pi_\theta(\hat{y}_{ij}\,|\,P_i)}{\pi_{\theta_{\text{old}}}(\hat{y}_{ij}\,|\,P_i)}$
be the importance weight,
which measures how much more likely the new policy is to generate $\hat{y}_{ij}$ compared to the old one. 
The GRPO surrogate loss is:
\begin{equation*}
\begin{aligned}
\mathcal{L}_{\text{GRPO}}(\theta)
= {} & \mathbb{E}_{i,j} \Big[
\min\big(r_{ij}(\theta)\,A_{ij},\;
\operatorname{clip}(\epsilon_c)\,A_{ij}\big)
\Big] \\
& {} - \beta\,\mathrm{KL}\!\left[\pi_{\theta_{\text{old}}}\,\|\,\pi_\theta\right]
\end{aligned}
\end{equation*}

\subsection{Curriculum Learning}
\label{sec:curriculum}
Our analysis in Section~\ref{sec:repo_gap} confirms that coordinating edits across multiple code hunks is a primary challenge for AVR models.
To address this, we adopt a curriculum learning strategy that organizes training from simple to complex repairs. 
This approach decouples the learning process by progressively increasing task complexity, allowing the model to master local security edits before tackling global syntactic synchronization.

We use the number of code hunks in human reference as a proxy for repair difficulty.
We define three stages: easy (1-2 hunks), medium (3-5 hunks), and hard (>5 hunks).
Preliminary experiments showed that training only on single-hunk fixes caused the model to overfit to overly localized repairs; therefore, our easy stage includes two-hunk functions.
Our curriculum proceeds through the three stages during the RL phase. The training set is \emph{expanded cumulatively} at each epoch to prevent catastrophic forgetting, and the model's policy is always initialized from the previous stage's checkpoint.


\section{Experimental Setup}
\subsection{Baselines and Evaluation Metrics}

We benchmark \sys against specialized AVR approaches (VulMaster~\cite{vulmaster}, FAVOR~\cite{favor}) retrained on our repository-level split, a commercial SOTA model (GPT-4o~\cite{achiam2023gpt}) prompted with identical instructions, and our own SFT-only base model (Qwen2.5-7B) to isolate the contributions of the RL stage. More details are in Appendix~\ref{app: baselines}.

\subsection{Datasets}
\label{sec: dataset}
To strictly prevent data leakage and evaluate cross-project generalization, we enforce a repository-level split where no repository overlaps between training, validation, and test sets. 
Following selected baselines, we utilize BigVul~\cite{fan2020ac}. In addition, we construct a new out-of-distribution test set PrimeVul\textsubscript{AVR} derived from the vulnerability detection dataset PrimeVul~\cite{primevul}. We rigorously filtered out any repositories present in the BigVul training set, resulting in 1,554 unique function pairs. Detailed construction pipelines and statistics are provided in Appendix~\ref{app:dataset_details}.



\subsection{Implementation Details}
We implement \sys based on Qwen2.5-7B-Instruct~\cite{bai2023qwen} using HuggingFace~\cite{transformers} and Verl~\cite{verl}. For SFT, we perform full-parameter fine-tuning for 3 epochs with a learning rate of $3\times 10^{-5}$ and a context length of 4,096, computing loss exclusively on LLM response. For RL, we employ the GRPO algorithm initialized from the best SFT checkpoint. We set the global batch size to 1,024 with $M=8$ rollouts per prompt and train for up to 20 epochs with an actor learning rate of $1\times 10^{-6}$. Full details are available at Appendix~\ref{app: implement_detail}.
We use CodeBLEU as our primary metric following existing works~\cite{vulmaster, favor}. 
We exclude the Exact Match (EM) metric from our main evaluation. Under the rigorous repository-level split, all models achieve negligible scores ($<5\%$), rendering the metric indistinguishable (more discussion of EM is in Appendix~\ref{app: em}).

\section{Results}

\begin{figure}[t]
    \centering
    \includegraphics[width=0.45\textwidth]{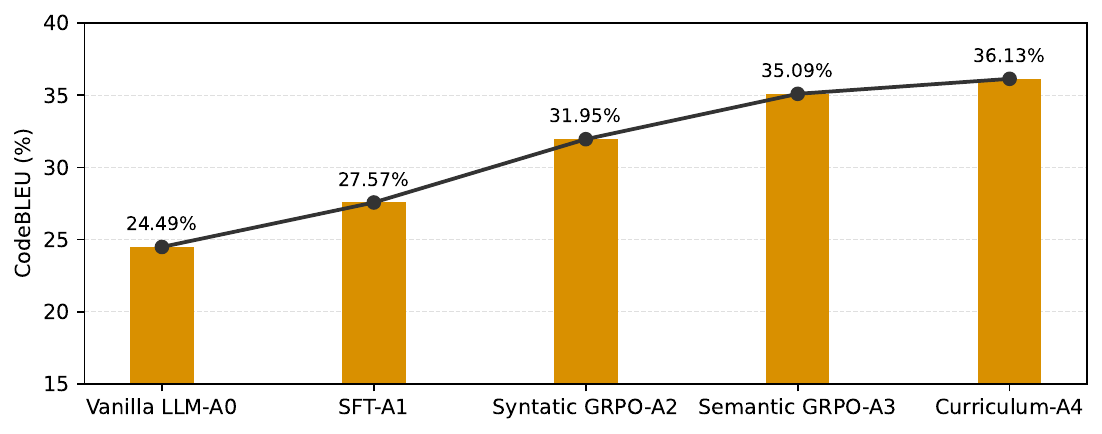}
    \caption{Step-wise ablation performance of \sys across training stages. We confirm that each technical design contributes to the final performance.}
    \label{fig:rq3_cmp3}
    \vspace{-5mm}
\end{figure}

\begin{figure}[t]
    \centering
    \includegraphics[width=0.45\textwidth]{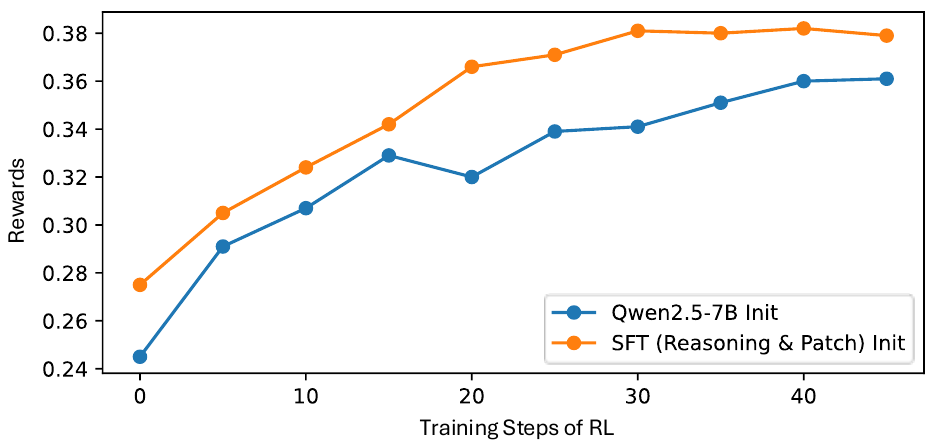}
    \vspace{-3mm}
    \caption{Training curves of the RL stage.}
    \vspace{-4mm}
    \label{fig:rq3_cmp2}
\end{figure}

\begin{figure}
     \centering
     \includegraphics[width=0.9\linewidth]{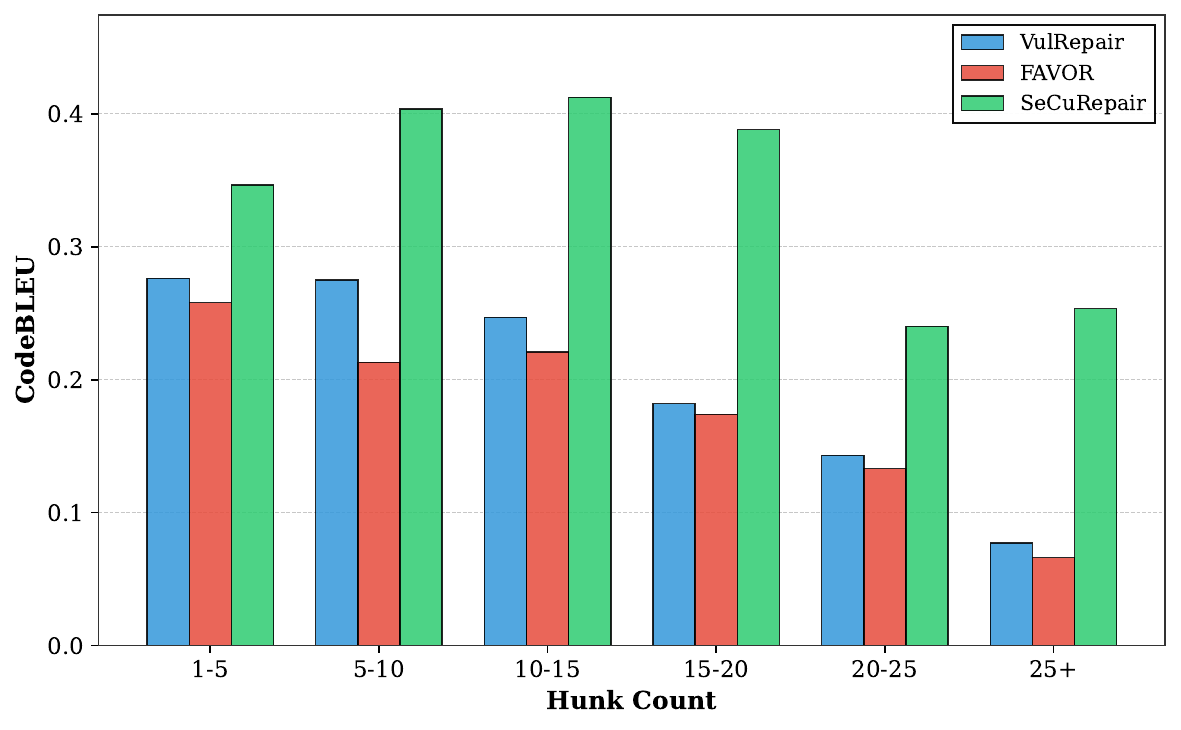}
     \caption{CodeBLEU across different hunk counts.}
     \label{fig:placeholder}
     \vspace{-5mm}
 \end{figure}

We evaluate the effectiveness of \sys in repairing vulnerabilities under both automatic and human evaluation. The automatic evaluation results are summarized in Table~\ref{tab:rq1_results}, with the human study results in Table~\ref{tab: human_preference}. We present the ablation performance of reasoning in Table~\ref{tab:rq3_cmp1} and Figure~\ref{fig:rq3_cmp2}, semantic-aware RL in Figure~\ref{fig:rq3_cmp3}, and curriculum training in Figure~\ref{fig:placeholder}.

\noindent\textbf{1) \sys performs better than SOTA baselines.}
Table~\ref{tab:rq1_results} shows that \sys outperforms all baselines on both datasets by a significant margin in terms of CodeBLEU.
Specifically, \sys surpasses the best-performing baseline on BigVul, VulMaster, by 37.21\%; and surpasses the best-performing baseline on PrimeVul\textsubscript{AVR}, GPT-4o, by 33.58\%.
This result demonstrates the superior ability of \sys to repair vulnerability and generalize to unseen repositories.
A Wilcoxon signed-rank test~\cite{woolson2007wilcoxon} confirms that all of \sys's performance gains over the baselines are statistically significant (p < 0.001).

Notablly, \sys outperforms the monolithic SFT baseline (Qwen2.5-7B+SFT) with CodeBLEU gains of 21.98\% on BigVul and 20.64\% on PrimeVul\textsubscript{AVR}, confirming the superiority of our training framework over standard SFT.

\noindent\textbf{2) Human evaluation validates the superior performance of \sys.} To complement automatic metrics, we conducted a blind human evaluation comparing \sys against VulMaster and GPT-4o. We recruited four experts with experience in software security and C++ to evaluate 309 statistically sampled instances (95\% confidence level). Participants rated patch semantic similarity on a 5-point Likert scale relative to the human reference (full detail is available at Appendix~\ref{app: humanevaluation}). Aligning with the goal of AI-assisted repair~\cite{takerngsaksiri2025human}, we categorize scores $\ge 3$ as ``workable drafts,'' indicating patches that capture sufficient logic to serve as valid starting points for developers. 


\begin{table}
\caption{Comparison of \sys with state-of-the-art baselines on BigVul and PrimeVul\textsubscript{AVR}.}
\vspace{-2mm}
\label{tab:rq1_results}
\small
\centering
\begin{tabular}{lccc}
\toprule
\multirow{2}{*}{\textbf{Approach}} & 
\multirow{2}{*}{\textbf{Strategy}} & 
\multicolumn{2}{c}{\textbf{CodeBLEU (\%)}} \\
\cmidrule(lr){3-4}
 &  & \textbf{BigVul} & \textbf{PrimeVul\textsubscript{AVR}} \\
\midrule
FAVOR & SFT & 25.78 & 11.77 \\
VulMaster & SFT & 26.33 & 11.62 \\
GPT-4o & NA & 25.90 & 23.41 \\
\sys & SFT & 29.62 & 25.92 \\
\sys & SFT \& RL & \textbf{36.13} & \textbf{31.27} \\
\bottomrule
\end{tabular}
\end{table}

\begin{table}
\caption{Human Preference Distribution across Different AVR Systems.}
\vspace{-2mm}
\label{tab:human_score_dist}
\centering
\small
\begin{tabular}{lccc}
\toprule
\textbf{Score} &
\textbf{SeCuRepair} &
\textbf{GPT-4o} &
\textbf{VulMaster} \\
\midrule
1   & 9   & 28  & 83 \\
2   & 121 & 127 & 164 \\
3   & 124 & 104 & 47 \\
4   & 24  & 31  & 12 \\
5   & 31  & 19  & 3 \\
\midrule
\textbf{AVG} & \textbf{2.83} & \textbf{2.63} & \textbf{1.99} \\
\bottomrule
\end{tabular}
\vspace{-3mm}
\label{tab: human_preference}

\end{table}

\begin{table}[t]
\caption{Repair effectiveness of \textit{patch-only} vs 
\textit{reasoning+patch} supervision on BigVul.}
\vspace{-2mm}
\label{tab:rq3_cmp1}
\centering
\small
\resizebox{\linewidth}{!}{
\begin{tabular}{lcc}
\toprule
\textbf{Variant} & \textbf{SFT Data} & \textbf{CodeBLEU (\%)} \\
\midrule
Qwen2.5-7B-Instruct & NA & 24.49  \\
Qwen2.5-7B-Instruct$_{SFT}$ & Patch-Only & 25.78  \\
Qwen2.5-7B-Instruct$_{SFT}$ & Reasoning+Patch & \textbf{27.57} ($\uparrow 6.94\%$) \\
\bottomrule
\end{tabular}
}
\end{table}

The human evaluation result in Table~\ref{tab: human_preference} confirms the superiority of our approach with moderate Kappa agreement.
\sys achieves the highest rate of workable patches (58.0\%), outperforming GPT-4o (50.0\%) and VulMaster (20.0\%). This result validates that a specialized training pipeline is more effective for AVR than relying solely on large-scale generalist models. Qualitative analysis further reveals that \sys frequently generates patches that are syntactically distinct yet semantically identical to human references, demonstrating its capacity to learn semantic repair logic rather than token-level mimicry. (see Appendix~\ref{app: case_study} for case study).
Meanwhile, patches rated as ``workable'', despite minor issues like over-defensive logic or verbose syntax, successfully capture the underlying fix semantics (see Appendix~\ref{app: quantitative} for detailed definition and criteria).

\noindent\textbf{3) Each technical design progressively contributes to the final performance.}  
\sys trains LLMs in a multi-stage process. Therefore, we conduct a progressive ablation study on each design to quantify their contributions, as shown in Figure~\ref{fig:rq3_cmp3}. Specifically, we compare the following model variants:
1) Base Model (A0). The original, pre-trained Qwen2.5-7B-Instruct without any fine-tuning.
2) SFT with Reasoning (A1). The base model was fine-tuned on distilled reasoning traces.
3) SFT → Syntactic-aware RL (A2). We start from our A1 checkpoint and then apply RL, but with a reward function based only on the BLEU score. 
4) Syntactic-aware RL → Syntactic- and Semantic-aware RL (A3). We enhance the A2 checkpoint by replacing the BLEU-only reward with our full semantics-aware reward function (BLEU + AST + DFG + CFG). 
5) Full \sys Model (A4). Finally, we add the curriculum learning schedule to the A3 checkpoint
We confirm that each design progressively contributes to the final performance, highlighting the effectiveness of each component and the synergistic nature of \sys.

\noindent\textbf{4) Reasoning-enhanced initialization is effective.} 
We investigate whether the reasoning-enhanced SFT stage provides a better starting point for reinforcement learning. We compare two RL training runs: one initialized from our SFT checkpoint ($\theta_{SFT}$), and another that starts directly from the base model. Figure~\ref{fig:rq3_cmp2} plots the learning curves for both settings. We observe SFT-initialized policy outperforms the non-SFT variant by exhibiting faster convergence (reaching peak reward 33\% earlier), greater stability (avoiding mid-training volatility), and a higher final reward plateau (0.38 vs. 0.36), which directly correlates with improved downstream repair performance.

\noindent\textbf{5) Semantic-aware reward improves repairing performance over lexical reward.}  
As shown in Figure~\ref{fig:rq3_cmp3}, GRPO with our semantic-aware reward performs better than using a lexical reward signal that compares the string similarity to the human reference based on BLEU(from A2 to A3 checkpoint). 
This reward design helps the model to learn from syntactically different but semantically similar patches to the human reference, guiding the model to learn the underlying repair logic over lexical mimicry, leading to 7.67\% relative improvement in CodeBLEU.

\noindent\textbf{6) Difficulty-aware curriculum adds performance boost.}  
 We observe from Fig~\ref{fig:rq3_cmp3} (from A3 to A4), introducing a curriculum learning schedule improves the performance of \sys by 2.97\%. A per-bucket analysis reveals that the curriculum yields substantial improvements on complex fix: functions with 2-10 code hunks by 4.24\% and functions with >10 vulnerable regions by 9.46\% in terms of CodeBLEU. 
 In addition, we compare the performance of \sys and baselines across different count hunk in Figure~\ref{fig:placeholder}, we observe \sys consistently outperforms baselines in all hunk counts, highlighting the effectiveness of this curriculum training strategy.

\section{Analysis}
\noindent\textbf{1) SFT with vs. without Reasoning.} 
We assess whether training on explicit reasoning traces improves patch quality. We compare two SFT Qwen-2.5-7B: one trained on our distilled (vulnerable code, reasoning, ground-truth patch) triples, and the other model trained on the same dataset without reasoning text, i.e., (vulnerable code, ground-truth patch). 
The results in Table~\ref{tab:rq3_cmp1} show that while standard SFT improves over the base model, incorporating reasoning provides a distinct advantage. SFT (Reasoning+Patch) outperforms SFT (Patch-Only) by a margin of 6.94\% on CodeBLEU.
This performance gain indicates that training on explicit reasoning helps the model learn to coordinate semantic edits, moving it beyond simple token-level imitation toward a more robust understanding of the repair task.

\noindent\textbf{2) Performance on Critical Vulnerabilities.} 
We further analyze performance across the MITRE Top-10 most dangerous CWEs~\cite{mostdangerouscwe}. As shown in Table~\ref{tab:top10_cwe}, \sys consistently outperforms the SFT baseline across nearly all categories. The gains are particularly pronounced for high-impact injection vulnerabilities: SQL Injection (CWE-89) improves from 0.201 to 0.430, and XSS (CWE-79) from 0.232 to 0.303. These results confirm that our semantics-aware RL is highly effective at capturing the precise structural patterns required to mitigate complex security flaws.
Full results are available at Appendix~\ref{app: cwe}.

\noindent\textbf{3) Robustness to Code Perturbation.} 
We perform an ablation study and confirm that SeCuRepair is robust to code perturbation. Following the existing code perturbation strategy~\cite{yang2022natural}, we replace the first local variable in the input with a randomized string to create a syntactically valid variant. Since local variable names do not dictate control flow, this modification guarantees that the code syntax remains strictly equivalent, with minimal affect to code semantics. On the perturbed PrimeVul dataset, VulMaster's CodeBLEU dropped by 14.75\%, while SeCuRepair declined only 3.39\%, highlighting the superior robustness of \sys to code perturbation.


\section{Conclusion}
We reveal that current AVR approaches struggle with syntactic brittleness and multi-hunk complexity. In response, we propose \sys, which leverages a reason-then-edit mechanism, semantics-driven RL, and curriculum-training strategy to ground repairs in diagnostic planning and structural correctness. \sys achieves state-of-the-art performance on BigVul and our proposed PrimeVul\textsubscript{AVR}, validating the necessity of aligning training objectives with code semantics rather than surface forms. We plan to broaden \sys to diverse programming languages in future work.

\section*{Limitations}

We propose SeCuRepair, an automated vulnerability repair framework that synergizes expert-level reasoning with semantics-aware reinforcement learning and curriculum-training. 
While SeCuRepair demonstrates significant improvements over state-of-the-art baselines, we acknowledge several primary limitations. 
First, reliance on static proxy metrics. Our training framework utilizes static semantic rewards rather than execution-based feedback. While execution feedback provides ground-truth verification, it is computationally expensive and brittle in practice. We intentionally relax this constraint to ensure scalability across large-scale, real-world datasets (e.g., BigVul) where build environments are often incomplete. Although strictly "noisier" than dynamic testing, our results confirm that these proxies effectively guide the model toward effective repair.
Meanwhile, this reward remains grounded in reference comparison. 
However, we posit that this approach serves as a middle ground to mitigate syntactic overfitting. It enables \sys to recognize and reward patches that are syntactically diverse yet structurally faithful to the developer's intent, marking a crucial step toward fully reference-agnostic semantic repair.
Second, language specificity. Our current evaluation is restricted to C/C++ vulnerabilities. Although the underlying reason-then-edit paradigm and semantic-aware RL are designed to be language-agnostic, their generalizability to other ecosystems (e.g., Python, Java) or multi-lingual contexts remains to be empirically validated. 
Third, dependency on closed-source teachers. Our cold-start reasoning data is distilled from proprietary LLMs (e.g., GPT-4o). The black-box nature of these teachers poses challenges for full reproducibility and may introduce distillation bias. To mitigate these issues, we release our distilled datasets and enforce a strict security-expert aligned workflow to filter low-quality rationales, ensuring the student model learns robust patterns.
Lastly, our automatic evaluation relies on static metric of CodeBLEU. To mitigate this issue, we complement with human evaluation, quativative and quantitative case studies, validating \sys's effectiveness to help developers.

\section*{Ethical Consideration}
The primary goal of SeCuRepair is to enhance software security by automating the remediation of software defects. We recognize the theoretical risk that such models could be repurposed for malicious vulnerability discovery. However, given the rapid growth of software vulnerabilities outpacing manual repair capacity, we argue that advancing automated defensive tools is an ethical imperative. All data utilized in this study are publicly available, and we strictly adhere to the usage policies of the respective data sources.

\bibliography{custom}

@misc{nvd_dashboard,
    title={NVD Dashboard},
    howpublished={\url{https://nvd.nist.gov/general/nvd-dashboard}},
    note={(Accessed 07/09/2025)}
}

@article{cve_fix_time2,
  title   = {2020 State of the Octoverse: Securing the World's Software},
  author  = {Nicole Forsgren and Bas Alberts and Kevin Backhouse and Grey Baker and Greg Cecarelli and Derek Jedamski and Scot Kelly and Clair Sullivan},
  year    = {2021},
  journal = {arXiv preprint arXiv: 2110.10246}
}

@article{favor,
  journal = {2025 IEEE International Conference on Software Analysis, Evolution and Reengineering (SANER)},
  pages   = {193-204},
  doi     = {10.1109/SANER64311.2025.00026},
  title   = {Enhancing Automated Vulnerability Repair Through Dependency Embedding and Pattern Store},
  year    = {2025},
  author  = {Qingao Dong and Yuanzhang Lin and Hailong Sun and Xiang Gao},
  url     = {https://ieeexplore.ieee.org/document/10992522}
}

@inproceedings{takerngsaksiri2025human,
  title={Human-in-the-loop software development agents},
  author={Takerngsaksiri, Wannita and Pasuksmit, Jirat and Thongtanunam, Patanamon and Tantithamthavorn, Chakkrit and Zhang, Ruixiong and Jiang, Fan and Li, Jing and Cook, Evan and Chen, Kun and Wu, Ming},
  booktitle={2025 IEEE/ACM 47th International Conference on Software Engineering: Software Engineering in Practice (ICSE-SEIP)},
  pages={342--352},
  year={2025},
  organization={IEEE}
}

@inproceedings{vulmaster,
  title={Out of sight, out of mind: Better automatic vulnerability repair by broadening input ranges and sources},
  author={Zhou, Xin and Kim, Kisub and Xu, Bowen and Han, DongGyun and Lo, David},
  booktitle={Proceedings of the IEEE/ACM 46th international conference on software engineering},
  pages={1--13},
  year={2024}
}

@article{vrepair,
  title={Neural transfer learning for repairing security vulnerabilities in c code},
  author={Chen, Zimin and Kommrusch, Steve and Monperrus, Martin},
  journal={IEEE Transactions on Software Engineering},
  volume={49},
  number={1},
  pages={147--165},
  year={2022},
  publisher={IEEE}
}

@article{wang2021codet5,
  title={Codet5: Identifier-aware unified pre-trained encoder-decoder models for code understanding and generation},
  author={Wang, Yue and Wang, Weishi and Joty, Shafiq and Hoi, Steven CH},
  journal={arXiv preprint arXiv:2109.00859},
  year={2021}
}

@inproceedings{fan2020ac,
  title={AC/C++ code vulnerability dataset with code changes and CVE summaries},
  author={Fan, Jiahao and Li, Yi and Wang, Shaohua and Nguyen, Tien N},
  booktitle={Proceedings of the 17th international conference on mining software repositories},
  pages={508--512},
  year={2020}
}

@misc{cveexample,
title = {CVE-2021-29513},
howpublished = {\url{https://nvd.nist.gov/vuln/detail/CVE-2021-29513}},
month = {},
year = {},
note = {(Accessed on 12/09/2025)}
}

@misc{cwe476,
title = {CWE-476},
howpublished = {\url{https://cwe.mitre.org/data/definitions/476.html}},
month = {},
year = {},
note = {(Accessed on 12/09/2025)}
}

@article{achiam2023gpt,
  title={Gpt-4 technical report},
  author={Achiam, Josh and Adler, Steven and Agarwal, Sandhini and Ahmad, Lama and Akkaya, Ilge and Aleman, Florencia Leoni and Almeida, Diogo and Altenschmidt, Janko and Altman, Sam and Anadkat, Shyamal and others},
  journal={arXiv preprint arXiv:2303.08774},
  year={2023}
}

@article{bai2023qwen,
  title={Qwen technical report},
  author={Bai, Jinze and Bai, Shuai and Chu, Yunfei and Cui, Zeyu and Dang, Kai and Deng, Xiaodong and Fan, Yang and Ge, Wenbin and Han, Yu and Huang, Fei and others},
  journal={arXiv preprint arXiv:2309.16609},
  year={2023}
}

@article{woolson2007wilcoxon,
  title={Wilcoxon signed-rank test},
  author={Woolson, Robert F},
  journal={Wiley encyclopedia of clinical trials},
  pages={1--3},
  year={2007},
  publisher={Wiley Online Library}
}

@article{shao2024deepseekmath,
  title={Deepseekmath: Pushing the limits of mathematical reasoning in open language models},
  author={Shao, Zhihong and Wang, Peiyi and Zhu, Qihao and Xu, Runxin and Song, Junxiao and Bi, Xiao and Zhang, Haowei and Zhang, Mingchuan and Li, YK and Wu, Yang and others},
  journal={arXiv preprint arXiv:2402.03300},
  year={2024}
}

@inproceedings{lin2007autopag,
  title={AutoPaG: towards automated software patch generation with source code root cause identification and repair},
  author={Lin, Zhiqiang and Jiang, Xuxian and Xu, Dongyan and Mao, Bing and Xie, Li},
  booktitle={Proceedings of the 2nd ACM symposium on Information, computer and communications security},
  pages={329--340},
  year={2007}
}

@inproceedings{zhang2022program,
  title={Program vulnerability repair via inductive inference},
  author={Zhang, Yuntong and Gao, Xiang and Duck, Gregory J and Roychoudhury, Abhik},
  booktitle={Proceedings of the 31st ACM SIGSOFT International Symposium on Software Testing and Analysis},
  pages={691--702},
  year={2022}
}

@article{gao2021beyond,
  title={Beyond tests: Program vulnerability repair via crash constraint extraction},
  author={Gao, Xiang and Wang, Bo and Duck, Gregory J and Ji, Ruyi and Xiong, Yingfei and Roychoudhury, Abhik},
  journal={ACM Transactions on Software Engineering and Methodology (TOSEM)},
  volume={30},
  number={2},
  pages={1--27},
  year={2021},
  publisher={ACM New York, NY, USA}
}

@inproceedings{rahman2024causal,
  author    = {Md Mahbubur Rahman and Ira Ceka and Chengzhi Mao and Saikat Chakraborty and Baishakhi Ray and Wei Le},
  title     = {Towards Causal Deep Learning for Vulnerability Detection},
  booktitle = {Proceedings of the 46th {IEEE/ACM} International Conference on Software Engineering, {ICSE} 2024, Lisbon, Portugal, April 14-20, 2024},
  pages     = {153:1-153:11},
  publisher = {ACM},
  year      = {2024},
  url       = {https://doi.org/10.1145/3597503.3639170},
  doi       = {10.1145/3597503.3639170},
  bibsource = {dblp computer science bibliography, https://dblp.org}
}

@article{weyssow2025r2vul0,
  title={R2Vul: Learning to Reason about Software Vulnerabilities with Reinforcement Learning and Structured Reasoning Distillation},
  author={Weyssow, Martin and Yang, Chengran and Chen, Junkai and Widyasari, Ratnadira and Zhang, Ting and Huang, Huihui and Nguyen, Huu Hung and Tun, Yan Naing and Bui, Tan and Li, Yikun and others},
  journal={arXiv preprint arXiv:2504.04699},
  year={2025}
}

@article{primevul,
  title   = {Vulnerability Detection with Code Language Models: How Far Are We?},
  author  = {Yangruibo Ding and Yanjun Fu and Omniyyah Ibrahim and Chawin Sitawarin and Xinyun Chen and Basel Alomair and David Wagner and Baishakhi Ray and Yizheng Chen},
  year    = {2024},
  journal = {arXiv preprint arXiv: 2403.18624}
}

@inproceedings{zhang2023learning,
  title={Learning to locate and describe vulnerabilities},
  author={Zhang, Jian and Liu, Shangqing and Wang, Xu and Li, Tianlin and Liu, Yang},
  booktitle={2023 38th IEEE/ACM International Conference on Automated Software Engineering (ASE)},
  pages={332--344},
  year={2023},
  organization={IEEE}
}

@inproceedings{zhang2020learning,
  title={Learning to handle exceptions},
  author={Zhang, Jian and Wang, Xu and Zhang, Hongyu and Sun, Hailong and Pu, Yanjun and Liu, Xudong},
  booktitle={Proceedings of the 35th IEEE/ACM International Conference on Automated Software Engineering},
  pages={29--41},
  year={2020}
}

@misc{mostdangerouscwe,
  title={Most Dangerous CWEs},
  howpublished={\url{https://cwe.mitre.org/top25/archive/2024/2024_cwe_top25.html}},
  note={(Accessed 07/09/2025)}
}

@misc{transformers,
  title={Transformers},
  howpublished={\url{https://huggingface.co/docs/transformers/en/index}},
  note={(Accessed 07/09/2025)}
}

@misc{vllm,
  title={VLLM},
  howpublished={\url{https://docs.vllm.ai/en/latest/usage/index.html}},
  note={(Accessed 07/09/2025)}
}

@misc{verl,
  title={Verl},
  howpublished={\url{https://verl.readthedocs.io/en/latest/index.html}},
  note={(Accessed 07/09/2025)}
}

@article{yang2024acecode,
  title={ACECode: A Reinforcement Learning Framework for Aligning Code Efficiency and Correctness in Code Language Models},
  author={Yang, Chengran and Kang, Hong Jin and Shi, Jieke and Lo, David},
  journal={arXiv preprint arXiv:2412.17264},
  year={2024}
}

@inproceedings{aggarwal2025nextcoder,
  title={NextCoder: Robust Adaptation of Code LMs to Diverse Code Edits},
  author={Aggarwal, Tushar and Singh, Swayam and Awasthi, Abhijeet and Kanade, Aditya and Natarajan, Nagarajan},
  booktitle={Forty-second International Conference on Machine Learning},
  year={2025}
}

@inproceedings{ke2025niodebugger,
  title={NIODebugger: A novel approach to repair non-idempotent-outcome tests with LLM-based agent},
  author={Ke, Kaiyao},
  booktitle={2025 IEEE/ACM 47th International Conference on Software Engineering (ICSE)},
  pages={762--762},
  year={2025},
  organization={IEEE Computer Society}
}

@inproceedings{mahbub2023explaining,
  title={Explaining software bugs leveraging code structures in neural machine translation},
  author={Mahbub, Parvez and Shuvo, Ohiduzzaman and Rahman, Mohammad Masudur},
  booktitle={2023 IEEE/ACM 45th International Conference on Software Engineering (ICSE)},
  pages={640--652},
  year={2023},
  organization={IEEE}
}

@inproceedings{zhang2020retrieval,
  title={Retrieval-based neural source code summarization},
  author={Zhang, Jian and Wang, Xu and Zhang, Hongyu and Sun, Hailong and Liu, Xudong},
  booktitle={Proceedings of the ACM/IEEE 42nd International Conference on Software Engineering},
  pages={1385--1397},
  year={2020}
}

@article{hui2024qwen2,
  title={Qwen2. 5-coder technical report},
  author={Hui, Binyuan and Yang, Jian and Cui, Zeyu and Yang, Jiaxi and Liu, Dayiheng and Zhang, Lei and Liu, Tianyu and Zhang, Jiajun and Yu, Bowen and Lu, Keming and others},
  journal={arXiv preprint arXiv:2409.12186},
  year={2024}
}

@article{guo2025deepseek,
  title={Deepseek-r1: Incentivizing reasoning capability in llms via reinforcement learning},
  author={Guo, Daya and Yang, Dejian and Zhang, Haowei and Song, Junxiao and Zhang, Ruoyu and Xu, Runxin and Zhu, Qihao and Ma, Shirong and Wang, Peiyi and Bi, Xiao and others},
  journal={arXiv preprint arXiv:2501.12948},
  year={2025}
}

@article{wei2025swe,
  title={Swe-rl: Advancing llm reasoning via reinforcement learning on open software evolution},
  author={Wei, Yuxiang and Duchenne, Olivier and Copet, Jade and Carbonneaux, Quentin and Zhang, Lingming and Fried, Daniel and Synnaeve, Gabriel and Singh, Rishabh and Wang, Sida I},
  journal={arXiv preprint arXiv:2502.18449},
  year={2025}
}

@article{dutta2024applying,
  title={Applying RLAIF for code generation with API-usage in lightweight LLMs},
  author={Dutta, Sujan and Mahinder, Sayantan and Anantha, Raviteja and Bandyopadhyay, Bortik},
  journal={arXiv preprint arXiv:2406.20060},
  year={2024}
}

@article{sghaier2025leveraging,
  title={Leveraging Reward Models for Guiding Code Review Comment Generation},
  author={Sghaier, Oussama Ben and Tufano, Rosalia and Bavota, Gabriele and Sahraoui, Houari},
  journal={arXiv preprint arXiv:2506.04464},
  year={2025}
}

@article{zhang2025benchmarking,
  title={Benchmarking Large Language Models for Multi-Language Software Vulnerability Detection},
  author={Zhang, Ting and Yang, Chengran and Su, Yindu and Weyssow, Martin and Nguyen, Hung and Bui, Tan and Kang, Hong Jin and Li, Yikun and Ouh, Eng Lieh and Shar, Lwin Khin and others},
  journal={arXiv preprint arXiv:2503.01449},
  year={2025}
}

@inproceedings{yang2022natural,
  title={Natural attack for pre-trained models of code},
  author={Yang, Zhou and Shi, Jieke and He, Junda and Lo, David},
  booktitle={Proceedings of the 44th International Conference on Software Engineering},
  pages={1482--1493},
  year={2022}
}

@inproceedings{curriculumlearning,
author = {Bengio, Yoshua and Louradour, J\'{e}r\^{o}me and Collobert, Ronan and Weston, Jason},
title = {Curriculum learning},
year = {2009},
booktitle = {Proceedings of the 26th Annual International Conference on Machine Learning},
}

@article{shervashidze2011weisfeiler,
  title={Weisfeiler-lehman graph kernels.},
  author={Shervashidze, Nino and Schweitzer, Pascal and Van Leeuwen, Erik Jan and Mehlhorn, Kurt and Borgwardt, Karsten M},
  journal={Journal of Machine Learning Research},
  volume={12},
  number={9},
  year={2011}
}

\appendix

\section{Appendix}
\label{sec:appendix_A}
\section{Variable Pertubation}
\label{app: pertubation}
To probe whether AVR approaches learn real semantic repair patterns or merely overfit to lexical cues, we test the robustness of baselines (VulMaster~\cite{vulmaster}, FAVOR~\cite{favor}, and VulMaster's base LLM CodeT5~\cite{wang2021codet5}) under a simple refactoring: renaming local variables while preserving code semantics.
We randomly selected 10 cases where baseline models (trained and evaluated under random splits) correctly generated a patch, 
then applied consistent variable renamings. 
For each instance, we manually renamed locally initialized variables to contextually appropriate alternatives (e.g., swapping len for size), ensuring the code remained natural and semantically equivalent, before re-evaluating the models.

\textbf{Observation}.
All studied models are highly sensitive to these minor changes, failing to generate functionally equivalent patches in 40–70\% of cases, based on manual inspection by the first author.
Figure~\ref{fig:repo_gap_qualitative} provides an example of this failure. 
In the original patch (a), VulMaster correctly inserts a safety guard to prevent a buffer over-read. However, after simply renaming the variable \texttt{plenbytes} to \texttt{nbytes} (b), the model fails. It does not generate the equivalent safety check but instead hallucinates an irrelevant function call, leaving the vulnerability unresolved.
This case study provides evidence that the model \emph{has not learned the underlying semantic rule}, i.e., ``the return value of the decoding function must be checked before proceeding.'' 
Instead, it learns a superficial correlation between the specific variable name \texttt{plenbytes} and the patch. 
While in limited scale, the consistent failures strongly motivates the need for an AVR learning framework that prioritizes syntactic and semantic understanding of repair patterns over simple lexical mimicry.

\section{Suboptimal on Multi-Hunk Repair}
\label{app:multi_hunk_repair}

\begin{figure*}[t]
  \centering
  \includegraphics[width=0.65\linewidth]{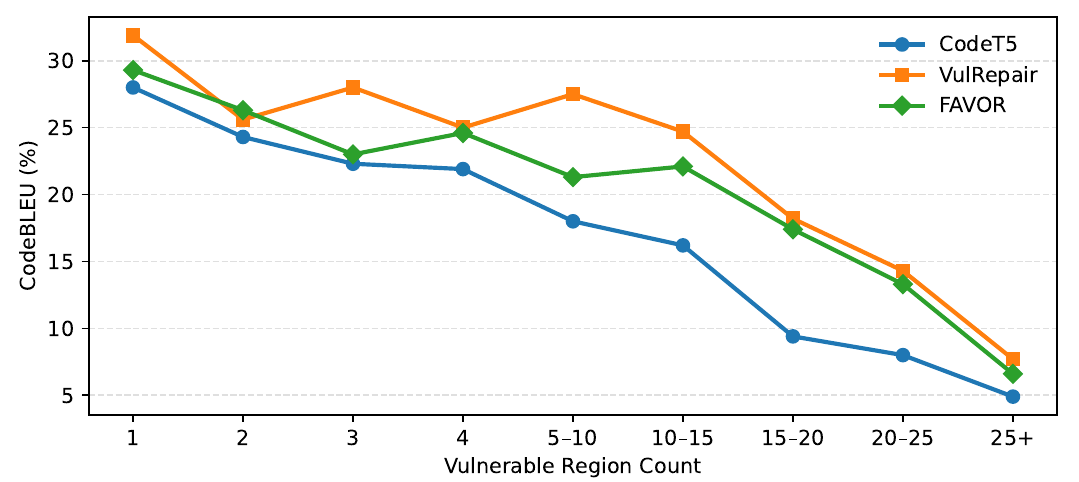} 
  \vspace{-3mm}
  \caption{The performance of existing AVR approaches drops as the number of hunks increases.}
  \label{fig:multi_hunk_repair}
  \vspace{-3mm}
\end{figure*}  

\noindent \textbf{Setup.}
Real-world vulnerability fixes frequently span multiple, non-contiguous regions (\emph{hunks}) that must be edited consistently. 
Prior work FAVOR~\cite{favor}'s case study notes that multi-hunk repair significantly increases AVR complexity and is a common failure mode for sequence-to-sequence models. 
To quantify this, we evaluate baseline models on our repository-split test set, stratifying the results by the number of hunks in each function. 

\textbf{Observation.} 
Figure~\ref{fig:multi_hunk_repair} shows that performance declines generally as the number of hunks increases. 
When moving from 1 to 2 hunks, VulMaster, FAVOR, and CodeT5 performance drops by 19.75\%, 11.27\%, and 13.22\% in terms of CodeBLEU, respectively.
This downward trend continues for repairs requiring more than two hunks.
This observation calls for an AVR approach that can work better in terms of multi-hunk repair performance.

\section{Cross-Repository Generalization Gap}
\label{app:repo_gap}

\begin{figure*}[t]
  \centering
  \includegraphics[width=0.95\linewidth]{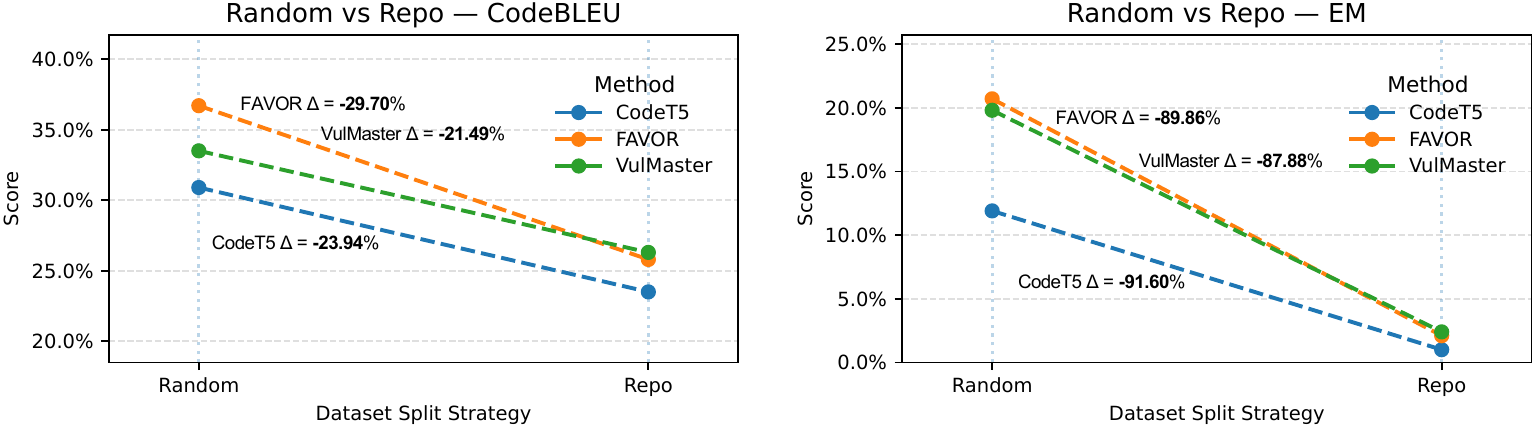} 
  \vspace{-2mm}
  \caption{Performance comparison of VulMaster, FAVOR, and CodeT5 under random split and repository-level split. 
  The performance of selected models degrades substantially under repository-level split by up to 91.6\% in terms of Exact Match and up to 29.7\% in terms of CodeBLEU. 
  }
  \label{fig:repo_gap}
  \vspace{-4mm}
\end{figure*}

\noindent \textbf{Setup.} 
AVR models must generalize beyond the repositories seen in training. 
We formalize this through a cross-repository evaluation protocol. 
We evaluate the baselines on the common dataset used by VulMaster and FAVOR's dataset: BigVul dataset~\cite{fan2020ac}. 
We compare their performance under two distinct data splitting strategies: the conventional random split (using the same split as FAVOR) and our strict repository-level split. 
In the repository-level split, all functions from a given project are confined to a single set (training, validation, or test).
To ensure a fair comparison, we fine-tune all models for each strategy. 
To prevent data contamination, we exclude the pre-trained adaptor modules of VulMaster, as their training relies on a bug-fixing corpus with known project-level overlaps with the BigVul test set. 

\textbf{Observation.} 
The results, summarized in Figure~\ref{fig:repo_gap}, reveal a significant cross-repository generalization gap. 
Under the strict repository-level split, the performance of all baseline models degrades sharply compared to the conventional random split. 
Across the tested models, CodeBLEU scores drop by a relative 21.49\% to 29.70\%, while EM scores plummet by 87.88\% to 89.86\%.
The collapse of the EM is particularly revealing. 
It strongly suggests that existing SFT-based AVR approaches are not learning semantic repair behaviors that can transfer to unseen projects. 
Instead, they are primarily overfitting on repository-specific surface forms and lexical patterns.

\section{Dataset}
\label{app:dataset_details}.
For fair comparison with prior AVR approaches, we adopt the BigVul corpus~\cite{fan2020ac} as used by VulMaster~\cite{vulmaster} and FAVOR~\cite{favor} for the main evaluation. 
We construct a \emph{repository-level} split: all functions from a given repository appear in exactly one of the training, validation, and test sets (no repository overlap).
We use an 8:1:1 train/val/test ratio and verify that no repository crosses splits.

To rigorously test how well models generalize to entirely unseen projects, we constructed PrimeVul\textsubscript{AVR}, a new AVR test dataset derived from PrimeVul~\cite{primevul}, a corpus originally created for vulnerability detection in C/C++.
Following the data extraction pipeline from VREPAIR~\cite{vrepair}, we created vulnerable-patched function pairs from PrimeVul's CVE-linked commits. To ensure a truly external test set, we then filtered out any function pair whose repository was present in our BigVul training set. This process resulted in a clean, out-of-distribution test set containing 1,554 C/C++ function pairs for evaluating cross-repository generalization.

\section{Implementation Details}
\label{app: implement_detail}
We implement \sys with HuggingFace Transformers~\cite{transformers} for SFT training, vllm~\cite{vllm} for inference, and Verl~\cite{verl} for RL training.
Due to excessive training costs for RL, we select one base model, Qwen2.5-7B-Instruct~\cite{bai2023qwen}, as proof-of-concept. 
Qwen2.5 is commonly used in software engineering tasks~\cite{yang2024acecode, aggarwal2025nextcoder, ke2025niodebugger} and relevant vulnerability detection task~\cite{weyssow2025r2vul0} as the base model.
This setting aligns with existing AVR approaches~\cite{vrepair,favor}, which fine-tune one base model.

\textbf{SFT implementation.}
We fine-tune Qwen2.5-7B-Instruct with full-parameter SFT.
Loss is computed only on model response: \texttt{<reason>} and \texttt{<patch>} spans.
Sequences are truncated to a cutoff length of 4096 tokens to avoid out-of-memory errors.
We train for 3 epochs with cosine learning rate decay and use 10\% of the dataset for warm-up.
Base learning rate is $3.0\times 10^{-5}$.
Per-device batch size is 4 with gradient accumulation of 4.

\textbf{RL implementation.}
We further optimize \sys with GRPO-style reinforcement learning using the Verl library, initialized from the best SFT checkpoint on the validation set.
We set the train batch size to 1024, with PPO mini-batches of 64.
The actor learning rate is \(1.0\times 10^{-6}\), with gradient checkpointing enabled and FSDP offloading for parameters and optimizer states to reduce memory overhead.
We generate M=8 rollouts per prompt.
We train for at most 20 epochs, saving checkpoints every epoch and saving the best checkpoint on the validation set.

For a controlled comparison, all open-source models (our baselines and \sys) are trained on our BigVul training split. During evaluation, we use deterministic decoding by setting the temperature to 0 for all models.

\section{Baselines}
\label{app: baselines}
We evaluate \sys against the following groups of baselines:
\begin{itemize}[leftmargin=*]
    \item \textbf{Learning-based AVR approaches}. We select two state-of-the-art AVR approaches, VulMaster~\cite{vulmaster} and FAVOR~\cite{favor}. 
    VulMaster applies CWE expert knowledge to guide the repair process and can handle long input sequences. 
    FAVOR augments the input function with CFG and historical patches. 
    We retrain both on our repository-level split using their recommended hyperparameters.
    \item \textbf{Commercial LLMs}. 
    We also evaluate against a top-tier commercial model GPT-4o~\cite{achiam2023gpt} to benchmark against the general-purpose state-of-the-art. We prompt GPT-4o with the same instructional wrapper used for \sys to ensure a fair, direct comparison of repair capabilities.
    \item \textbf{\sys base model with SFT}. 
    We fine-tune the base model of \sys, Qwen2.5-7B-Instruct~\cite{bai2023qwen}, on our repository-level BigVul training split as one baseline. We select the best checkpoint based on the performance of the validation set. 
\end{itemize}

\section{The Brittleness of the Exact Match Metric.}
\label{app: em}
The case study in Figure~\ref{fig:repo_gap_qualitative} highlights the fundamental brittleness of the Exact Match (EM) metric for evaluating AVR. 
EM is fundamentally sensitive to syntactic variations that preserve the code's semantics.
For example, a logically equivalent change like rewriting \texttt{if (a > b)} to \texttt{if (b < a)} would still result in an EM score of 0.
Indeed, an ideal AVR approach capable of semantic repairing \emph{should} be able to generate a diverse set of semantically correct and equivalent patches, most of which would be wrongly evaluated by EM as they are lexically different from the single oracle patch.
This strictness makes EM an unreliable indicator of a patch's correctness and ill-suited for evaluating advanced, semantics-aware AVR approaches. 
In contrast, CodeBLEU considers AST similarity, making it robust to syntactic variations. 
Therefore, we consider CodeBLEU a more meaningful metric for AVR and use it as our primary measure in the following sections.

\section{Human Evaluation.}
\label{app: humanevaluation}
\subsection{Experiment Setting.}
To complement automatic metrics, we conduct a human evaluation to assess the correctness of the generated patches. 
We compare \sys against the best-performing open-source (VulMaster) and commercial baseline (GPT-4o).
We recruited four participants, all of whom have at least two years of experience in software security and C/C++ programming. From the PrimeVul\textsubscript{AVR} test set, we randomly sampled 309 examples (calculated for a 95\% confidence level with a 5\% margin of error).
For each sample, participants were shown the vulnerable code alongside the generated patch (presented as a diff). The ground-truth patch was also provided for reference. To prevent bias, the outputs from the different models for each sample were presented in a random order.
Given the time-intensive nature of formally verifying a patch's functional correctness, we adopted a widely used proxy for correctness: participants evaluated the extent to which a generated patch preserved the same semantic functionality as the ground-truth solution. 
Following existing works~\cite{mahbub2023explaining, zhang2020retrieval, zhang2023learning,zhang2020learning}, each sample is rated by all four participants on a 5-point Likert scale (from 1: not similar at all to 5: exactly the same semantics).

In this work, we aim to support developers with useful repair suggestions rather than replace them with fully automated fixes. Accordingly, we treat scores 1–2 as negative, indicating that the generated patch is of poor quality and unsuitable as a draft for developers. Scores 3–5 are considered positive (workable patch drafts): a score of 3 reflects a patch that captures key logical elements of the oracle patch but is incomplete, while scores 4 and 5 correspond to nearly correct and fully correct patches, respectively. Thus, patches with scores $\ge$3 are regarded as workable, aligning with our goal of providing developers with actionable repair suggestions.  
1) Score-3: semantically similar but partial fixes, either incomplete or containing superfluous logic. 
2) Score-4: high semantic alignment but differing implementation. 
3) Score 5: exact logical match. 

Recent work~\cite{takerngsaksiri2025human} also emphasizes the importance of keeping humans in the loop, with automated tools serving primarily as recommenders. Following this perspective, we evaluate how well \sys delivers workable patch drafts that serve as strong starting points for developers.

\subsection{Case Analysis}
\label{app: case_study}

\begin{figure*}[t]
  \centering
  \includegraphics[width=0.98\textwidth]{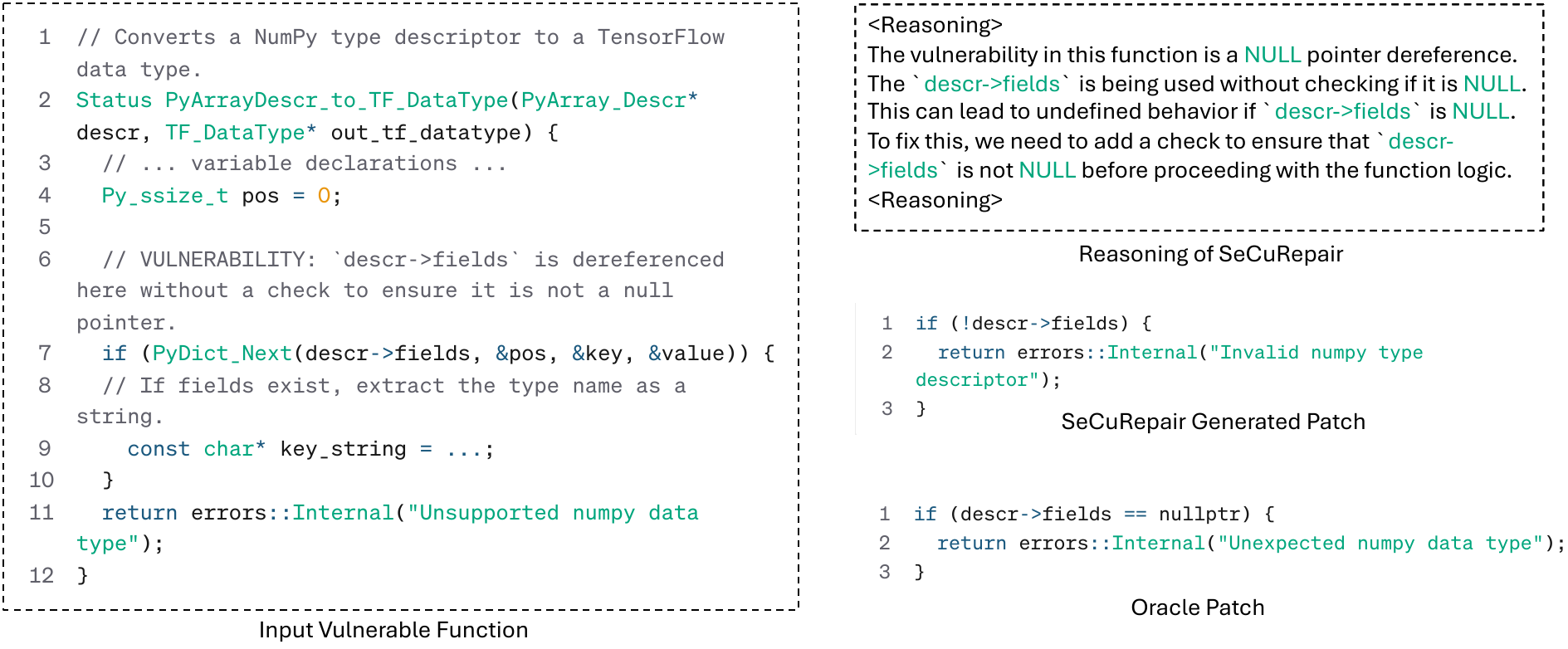}
  \caption{An example of \sys performing semantic repair with syntactic different patch with oracle.}
  \vspace{-3mm}
  \label{fig:case_study}
  \vspace{-3mm}
\end{figure*}

We show one example of \sys, in which \sys-generated patch is syntactically different but semantically identical to oracle, in Figure~\ref{fig:case_study}. This example highlights \sys's performance in terms of its reasoning quality, the generated patch's correctness, and its ability to achieve semantic equivalence through a syntactically different solution.

The vulnerable function sampled from PrimeVul\textsubscript{AVR} (\texttt{CVE-2021-29513}~\cite{cveexample}), contains a critical NULL pointer dereference vulnerability (CWE-476~\cite{cwe476}). The code at line 7 directly dereferences the \texttt{descr->fields} pointer in a call to \texttt{PyDict\_Next} without first verifying that it is not a null pointer. This could lead to a crash if the function is called with an improperly initialized descriptor.

As shown in Figure~\ref{fig:case_study}, \sys's reasoning for the fix is both concise and accurate. 
It correctly pinpoints the exact cause: the \texttt{descr->fields} pointer being used without validation. 
Furthermore, the generated patch is functionally correct and semantically equivalent to the oracle, successfully mitigating the vulnerability by introducing a guard condition before the pointer is dereferenced.

We also highlight that \sys's patch and oracle patch are syntactically different. 
Our system employs the C-style null check \texttt{!descr->fields}, while the oracle patch uses the explicit C++ \texttt{descr->fields == nullptr}. 
Additionally, the error messages, while functionally similar, use different strings. 
This distinction is significant as it reveals that the \sys is not merely performing a surface-level pattern match. 
Instead, it understands the underlying intent of the security fix, allowing it to generate a valid and effective repair. 

\subsection{Quantitative Analysis}
\label{app: quantitative}
Human annotators are asked to summarize the common failure modes in those workable but imperfect patches (scored 3 and 4). We observe score-3 patches exhibit three common failure modes: over-defensive fixes (e.g., excessive null checks that may alter intended behavior), minor logic gaps or partially applied fixes, and patches entangled with faulty refactoring. Score-4 patches are semantically similar to the Oracle but differ in implementation, often being syntactically verbose, less readable, over-commented, or using alternative APIs that may have minor semantics differences.

\section{Discussion of the most dangerous CWE performance.} 
\label{app: cwe}

To better understand the impact of \sys, we conducted a fine-grained analysis of model performance on the top-10 most dangerous CWE types~\cite{mostdangerouscwe} within the PrimeVul\textsubscript{AVR} dataset. Table \ref{tab:top10_cwe} compares the CodeBLEU scores of the full \sys framework against \sys-SFT, the SFT-only variant of \sys that performs best among all baselines.

The results clearly demonstrate the benefits of \sys. \sys consistently outperforms \sys-SFT across nearly all the most dangerous CWE categories where data is available. 
The improvements are particularly pronounced for high-impact vulnerabilities such as SQL Injection (CWE-89), where \sys achieves a score of 0.430 compared to 0.201 for \sys-SFT, and Cross-Site Scripting (CWE-79), with an improvement from 0.232 to 0.303. 
This suggests that the RL stage with syntactic- and semantics-aware reward is highly effective at guiding the model to learn the structural patterns required to fix common injection and memory safety flaws.

\begin{table*} [t]
\vspace{-3mm}
\caption{Comparison of \sys with \sys-SFT on top-10 most dangerous CWE.}
\vspace{-3mm}
\label{tab:top10_cwe}
\resizebox{\textwidth}{!}{
\begin{tabular}{llcccc}
\toprule
Rank & CWE-Type & Name & \sys & \sys-SFT & \# Samples \\
\midrule
1 & CWE-79 & Cross-Site Scripting & \textbf{0.303} & 0.232 & 7 \\
2 & CWE-787 & Out-of-bounds Write & \textbf{0.313} & 0.273 & 230 \\
3 & CWE-89 & SQL Injection & \textbf{0.430} & 0.201 & 2 \\
4 & CWE-352 & Cross-Site Request Forgery & \textbf{0.541} & 0.486 & 1 \\
5 & CWE-22 & Path Traversal & \textbf{0.297} & 0.267 & 23 \\
6 & CWE-125 & Out-of-bounds Read & \textbf{0.359} & 0.276 & 174 \\
7 & CWE-78 & OS Command Injection & 0.160 & \textbf{0.165} & 9 \\
8 & CWE-416 & Use After Free & \textbf{0.272} & 0.265 & 57 \\
9 & CWE-862 & Missing Authorization & NA & NA & 0 \\
10 & CWE-434 & Unrestricted Upload of File with Dangerous Type & NA & NA & 0 \\
\midrule
\end{tabular}
}
\vspace{-4mm}
\end{table*}

\end{document}